\documentclass{easychair} % this is for ISCA conference proceedings

\usepackage{doc}
\usepackage{tikz}
\usetikzlibrary{shapes,backgrounds}
\usepackage{amsmath,amsfonts,amsthm}
\usepackage{tabularx}
\usepackage{booktabs}
\newcolumntype{Y}{>{\centering\arraybackslash}X}
\usepackage{stmaryrd}
    \usepackage{subcaption}
    
\title{Identification of Deregulated Transcription Factors\\
 Involved in Specific Bladder Cancer Subtypes
}

\author{
Magali Champion \inst{1}
\and
   Julien Chiquet\inst{2} 
\and
  Pierre Neuvial\inst{3}
  \and 
  Mohamed Elati \inst{4}
  \and \\
  François Radvanyi \inst{5}
  \and 
  Etienne Birmelé \inst{1}
}

\institute{
  Laboratoire MAP5, UMR8145 CNRS, Université Paris Descartes,
  Paris, France\\
  \email{magali.champion@parisdescartes.fr}
\and
   UMR MIA-Paris, AgroParisTech, INRA, Université Paris-Saclay, Paris, France\\
\and
Institut de Mathématiques de Toulouse, UMR 5219, Université de Toulouse, CNRS, France\\
\and
   Université de Lille, INSERM U908, Cell Plasticity and Cancer, Lille, France
\and
Institut Curie, PSL Research University, CNRS, UMR144, Paris, France
 }

\authorrunning{Champion, Chiquet, Neuvial, Elati, Radvanyi, Birmelé}

\titlerunning{Identification of Deregulated TFs in Bladder Cancer}

\begin{document}

\maketitle

\begin{abstract}
 Comparison between tumoral and healthy cells may reveal abnormal regulation behaviors between a transcription factor and the genes it regulates, without exhibiting differential expression of the former genes.  We propose a methodology for the identification of transcription factors involved in the deregulation of genes in tumoral cells. This strategy is based on the inference of a reference gene regulatory network that connects transcription factors to their downstream targets using gene expression data. Gene expression levels in tumor samples are then carefully compared to this reference network to detect deregulated target genes. A linear model is finally used to measure the ability of each transcription factor to explain these deregulations.
We assess the performance of our method by numerical experiments on a public bladder cancer data set derived from the Cancer Genome Atlas project. We identify genes known for their implication in the development of specific bladder cancer subtypes as well as new potential biomarkers.
\end{abstract}

%------------------------------------------------------------------------------
\section{Introduction}
\label{sect:introduction}

Today, after decades of intensive research, cancer is still one of the most deadly diseases worldwide, killing millions of people every year. Cancer is mainly caused by somatic mutations that affect critical genes and pathways. These mutations are mostly triggered by environmental factors (e.g. obesity, smoking, alcohol, lifestyle,...) often promoted by certain genetic configurations.
% To cure it, past research mostly focused on the effect of these environmental factors \citep{Tuyns79,Doll81} but, more recently, also on internal factors (e.g. inherited mutations, copy number alterations, hypo/hyper-methylation, over/under-expression,...) \citep{Perou00,Shlien09,Kulis10}.
 In the last two decades, large-scale projects, such as the Cancer Genome Atlas project (TCGA), which has produced massive amounts of multi-omics data, have launched to improve our understanding of cancers \cite{TCGA2}. In this context, developing statistical algorithms able to interpret these large data sets and to identify genes that are the origin of diseases and their causal pathways still remains an important challenge.

Genes are commonly affected by genomic changes in the pathogenesis of human cancer. Cancer is moreover a heterogeneous disease, with affected gene sets that may be highly different depending on subtypes, and thus requires different treatments of patients. Specific analyses of subtypes have for example revealed significant differences between breast cancer subgroups \cite{Lehman11} but also pancancer similarities between breast and bladder cancer subgroups \cite{Damrauer14}. 

Using transcriptional data allows to look beyond DNA, that is to study abnormalities in terms of gene expression. As a common approach, differential expression analysis, for which statistical procedures have been intensively explored, can be performed and altered genes are then differentially expressed genes
\cite{Kaczkowski16}.
 This points to relevant genes but does not take into account the regulations (activation and inhibition) between genes.
 %, which we consider as crucial in the notion of deregulation. 
 
The approach we consider consists in taking into account the regulation structure between genes. We particularly focus on transcription factors (TFs), for their major role played in the regulation of gene expression, which make them an attractive target for cancer therapy
% and have been extensively studied, especially as targets for cancer therapy
 \cite{Nebert02,Yeh13}. 
%Indeed, TFs play a preponderant role in the regulation of gene expression: by binding the promoter region of their target genes, TFs can activate or inhibit their expression, which make them an attractive target for cancer therapy \cite{Yeh13}. 
Regulation processes between TFs and their targets are usually represented by Gene Regulatory Networks (GRNs).
%, which give an overview of the mechanisms of cancer. 
In the last few years, many different methods have been proposed to infer GRNs %inference problem 
from collections of gene expression data. In a discrete framework, gene expression can be discretized depending on their status (under/over-expressed or normal) and truth tables provide the regulation structure \cite{Elati11}. In the continuous case, regression methods, including the popular Lasso \cite{Tibshirani96} and its derivatives, have provided powerful results \cite{Vignes11,Liu08}.

A deregulated gene then corresponds to a gene whose expression does not correspond to the expression level expected from its regulators expression. 
It is different from the notion of differential expression since a loss of regulation between a target gene and one of its regulating TFs implies a loss of correlation between them but not necessarily differential expression.
Conversely, a TF can be differentially expressed and one of its targets not, precisely because it is deregulated.
%This notion is different from the notion of differential expression. Indeed, consider a target gene regulated by a single TF which activates it. A loss of that regulation, for instance due to a mutation in the TF, will imply a loss in the correlation between the two genes but may not imply diffential expression. A contrario, one ca imagine a scenario where the TF is differentally expressed and the target gene is not, precisely because it is deregulated. 

To discover deregulated genes, a first possibility is to infer one network per condition and to compare them. Statistical difficulties due to the noisy nature of transcriptomic data and the large number of features compared to the sample size can be taken into account by inferring the networks jointly and penalizing the differences between them \cite{Chiquet11}. A second possible approach is to assess the adequacy of gene expression in tumoral cells to a reference GRN, in order to exhibit the most striking discrepancies, i.e. the regulations which are not fulfilled by the data \cite{Guziolowski09,Karlebach12}. Such methods however focus on checking the validity of the network rather than highlighting genes with an abnormal behavior. Finally, analyses may be conducted at the pathway level rather than the gene level \cite{Tarca09,Vaske10}. They are then not network-wide in the sense that each gene has a deregulation score by pathway it belongs to and pathways are treated independently. Moreover, as the pathways are extracted from curated databases, the regulations taken into account are not tissue-specific.

Here, we propose a statistical deregulation model that uses gene expression data to identify deregulated TFs involved in specific subtypes of cancer. This paper is organized as follows:
in Section 1, we present the 3-steps method we developed and our validation procedure. In Section 2, we illustrate its interest on the TCGA bladder cancer data set. We show that it can be used complementary to differential expression analysis to point to potential biomarkers of cancers.

%------------------------------------------------------------------------------
\vspace{-0.1cm}
\section{Methods}\label{sec:methods}
\subsection{Overview of the Procedure}
\begin{figure}[!ht]
\begin{center}
\includegraphics[width=14cm]{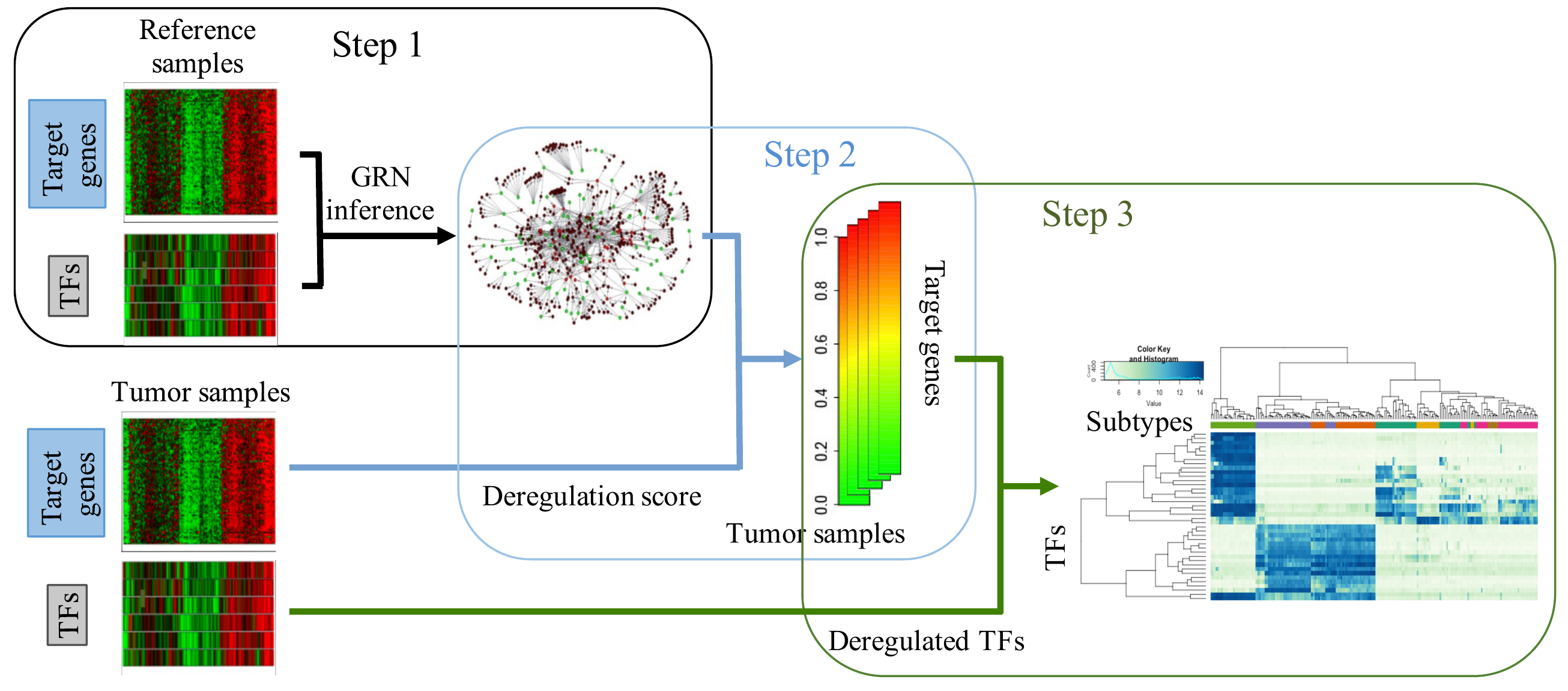}
\end{center}
\caption{Workflow of the proposed 3-steps algorithm for identifying TFs involved in specific cancer subtypes. 
%This algorithm is based on gene expression measured on both tumor and reference tissues for both transcription factors and target genes.
}\label{fig:overview}
\end{figure}

Our approach for the identification of deregulated transcription factors (TFs) involved in
cancers

\noindent is based on a 3-steps strategy that $(i)$ creates a reference gene regulatory network (GRN), which represents regulations between groups of co-expressed TFs and target genes using a reference data set (Step 1), $(ii)$ computes a deregulation score for each target gene in each tumor sample by comparing their behavior with the reference GRN (Step 2), $(iii)$ identifies the most significant TFs involved in the deregulation of the target genes in each sample from specific cancer subtypes (Step 3).
These steps are presented in Figure \ref{fig:overview} and described in detail in the next sections.

%------------------------------------------------------------------------------
\vspace{-0.1cm}
\subsection{Step 1: Inferring a Gene Regulatory Network}
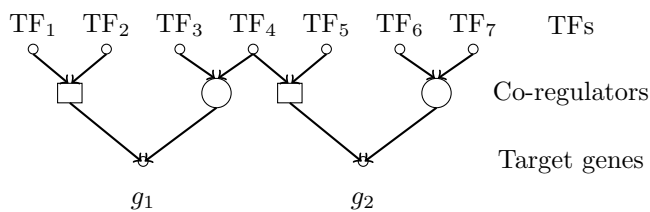
\begin{figure}[!ht]
\begin{center}
\begin{tikzpicture}[scale=0.65]
\draw (0,0) node{TF$_1$};
\draw (0,-0.5) circle(0.1);
\draw (1.5,0) node{TF$_2$};
\draw (1.5,-0.5) circle(0.1);
\draw (3,0) node{TF$_3$};
\draw (3,-0.5) circle(0.1);
\draw (4.5,0) node{TF$_4$};
\draw (4.5,-0.5) circle(0.1);
\draw (6,0) node{TF$_5$};
\draw (6,-0.5) circle(0.1);
\draw (7.5,0) node{TF$_6$};
\draw (7.5,-0.5) circle(0.1);
\draw (9,0) node{TF$_7$};
\draw (9,-0.5) circle(0.1);

\draw (0.5,-1.2) rectangle (1,-1.6);
\draw (3.75,-1.4) circle(0.3);
\draw (5,-1.2) rectangle (5.5,-1.6);
\draw (8.25,-1.4) circle(0.3);

\draw[->,thick] (0,-0.6) -- (0.72,-1.2);
\draw[->,thick] (1.5,-0.6) -- (0.78,-1.2);
\draw[->,thick] (3,-0.6) -- (3.72,-1.1);
\draw[->,thick](4.5,-0.6) -- (3.78,-1.1);
\draw[->,thick] (4.5,-0.6)--(5.22,-1.2);
\draw[->,thick](6,-0.6) -- (5.28,-1.2);
\draw[->,thick] (7.5,-0.6)--(8.22,-1.1);
\draw[->,thick] (9,-0.6)--(8.28,-1.1);

\draw (2.25,-3.6) node{$g_1$};
\draw (6.75,-3.6) node{$g_2$};

\draw (2.25,-2.8) circle(0.1);
\draw (6.75,-2.8) circle(0.1);

\draw[->,thick] (0.75,-1.6)--(2.22,-2.8);
\draw[->,thick] (3.75,-1.7)--(2.28,-2.8);
\draw[->,thick] (5.25,-1.6)--(6.72,-2.8);
\draw[->,thick] (8.25,-1.7)--(6.78,-2.8);

\draw (11,0) node{TFs};
%\draw (11,-0.5) node{factors};
\draw (11,-1.4) node{Co-regulators};
\draw (11,-2.8) node{Target genes};
\end{tikzpicture}
\end{center}
\caption{Example of LICORN graph involving 7 TFs and 2 target genes. TFs are gathered into groups of co-expressed genes that co-regulate (square for co-activators, circle for co-inhibitors) each target gene.
%: gene $g_1$ is activated by (TF$_1$,TF$_2$) and inhibited by (TF$_3$,TF$_4$). TF$_4$ activates $g_2$ while inhibiting $g_1$.
}\label{fig:licorn}
\end{figure}

Step 1 of the algorithm consists in inferring a GRN that connects TFs to their downstream targets. Among the large number of existing methods, we choose hLICORN, available in the {\tt CoRegNet} R-package \cite{CoRegNet}. This algorithm is based on a hybrid version of the LICORN model \cite{Elati07}, in which groups of co-regulated TFs act together to regulate the expression of their targets (Figure \ref{fig:licorn}). More precisely, LICORN uses heuristic techniques to identify co-activator and co-inhibitor sets from discretized gene expression matrices and locally associates each target gene to pairs of co-activators and co-inhibitors that significantly explain its discretized expression.
The hybrid variation of LICORN then ranks the local candidate networks according to how well they predict the target gene expression, through a linear regression, and selects the GRN that minimizes the prediction error.
This selection step limits the effects of overfitting, induced by the model complexity, especially the large number of features (genes) as compared to the sample size \cite{Chebil14}.
In this work, we slightly enrich the LICORN model by creating a copy of each TF in the target layer to allow regulations between TFs.

%, such allowing to infer the regulation structure controlling a given TF. This view is clearly a simplification of reality as the regulator and regulated roles of a TF  are considered independent. However, all GRN models are simplifications of the real biological mechanisms and may however allow to point out relevant TFs through well-chosen measures, as for instance the {\em influence} measure introduced in \citet{CoRegNet}. 

To construct a specific GRN, note that one may prefer using another inference method \cite{Chiquet12} or a pre-existing regulatory network, which can be loaded from the RegNetwork database \cite{Liu15}.
Here, we focus on hLICORN since the induced model is particularly suitable for the rest of our analysis. In addition, it was shown to provide powerful results for cooperative regulation detection, especially on cancer data set  \cite{Elati07,CoRegNet}. 

%------------------------------------------------------------------------------
\vspace{-0.1cm}
\subsection{Step 2: Computing a Deregulation Score}\label{sec-EM}
Step 2 of the algorithm aims at identifying deregulated target genes by carefully comparing their expression across all tumor samples with the reference GRN inferred in Step 1. For this purpose, we use the method described in \cite{deregScore}, which assumes that all genes from a hLICORN model are allowed to be deregulated, i.e. not to respond to their regulators as expected. 

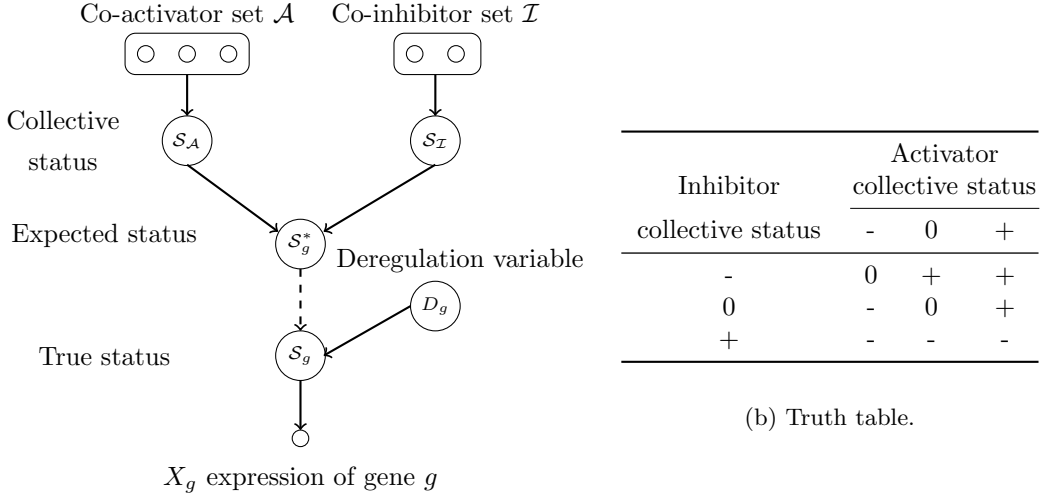
\begin{figure}[!ht]
\begin{subfigure}{0.53\textwidth}
%\begin{figure}[!ht]
\begin{center}
\begin{tikzpicture}[scale=1.1]
\draw (0,0) circle(0.1);
\draw (0.5,0) circle(0.1);
\draw (1,0) circle(0.1);
\draw (3.25,0) circle(0.1);
\draw (3.75,0) circle(0.1);

\draw[rounded corners] (-0.25,0.25) rectangle (1.25,-0.25); 
\draw[rounded corners] (3,0.25) rectangle (4.05,-0.25); 
\draw (0.5,0.5) node{Co-activator set $\mathcal{A}$};
\draw (3.5,0.5) node{Co-inhibitor set $\mathcal{I}$};

\draw[->,thick] (0.5,-0.25)--(0.5,-0.75);
\draw[->,thick] (3.5,-0.25)--(3.5,-0.75);
\draw(0.5,-1.05) circle(0.3);
\draw(3.5,-1.05) circle(0.3);
\draw (0.5,-1.05) node{\scriptsize $\mathcal{S}_{\mathcal{A}}$};
\draw (3.5,-1.05) node{\scriptsize $\mathcal{S}_{\mathcal{I}}$};
\draw (-1,-0.8) node{Collective};
\draw (-1,-1.3) node{status};
\draw (1.87,-2.3) node{\scriptsize $\mathcal{S}^*_g$};
\draw (1.87,-2.3) circle(0.3);
\draw[->,thick] (0.5,-1.34)--(1.60,-2.15);
\draw[->,thick] (3.5,-1.34)--(2.15,-2.15);
\draw (-0.5,-2.2) node{Expected status};
\draw (1.87,-3.65) node{\scriptsize $\mathcal{S}_g$};
\draw (1.87,-3.65) circle(0.3);
\draw (-0.5,-3.65) node{True status};
\draw[->,thick,dashed] (1.87,-2.6)--(1.87,-3.35);

\draw (1.87,-4.65) circle(0.1);
\draw (1.87,-5.15) node{ $X_g$ expression of gene $g$};
\draw[->,thick] (1.87,-3.95)--(1.87,-4.55);

\draw (3.5,-3.05) node{\scriptsize $D_g$};
\draw (3.5,-3.05) circle(0.3);
\draw (3.8,-2.5) node{Deregulation variable};
\draw[->,thick] (3.2,-3.05)--(2.15,-3.65);
\end{tikzpicture}
\end{center}
\caption{Deregulation model.}
\end{subfigure}
\begin{subfigure}{0.45\textwidth}
%\begin{table}[!ht]
\begin{center}
 \begin{tabularx}{0.85\textwidth}{c c c c}
\toprule
&  \multicolumn{3}{c}{Activator}  \\
 Inhibitor & \multicolumn{3}{c}{collective status}  \\
\cmidrule(lr){2-4} 
  collective status & - & 0 & + \\
\midrule
 -  & \textcolor{white}{e}0\textcolor{white}{e} &  + &  + \\
 0  &  - & 0 &  + \\
 +  &  - & - & - \\
\bottomrule
\end{tabularx}
\end{center}
%\caption{LICORN truth table, which gives the expected status of a target gene according to the collective status of its co-activators and co-inhibitors. The collective status are set by default to 0 except if and only if all of its elements share the same status. This table was established by biological considerations \cite{Elati07}.}\label{table:truth}
%\end{table}
\caption{Truth table.}
\end{subfigure}
\caption{(a) The deregulation model \cite{deregScore} used to compute a deregulation score for each target gene in each sample: each gene $g$ is associated to a hidden status $\mathcal{S}_g$ (under, over-expressed or normal). Target genes are allowed to be deregulated, i.e. not follow their co-regulator rules (Truth table (b)). The binary variable $D_g$ indicates whether the corresponding target gene $g$ is deregulated ($D_g=1$) or not ($D_g=0$). The deregulation score $Y$ of gene $g$ in sample $j$ is then the probability, given the observation, that $D_g=1$ in sample $j$.
(b) LICORN truth table, which gives the expected status of a target gene according to the collective status of its co-activators and co-inhibitors. Collective status are set by default to 0 except if and only if all of its elements share the same status. This table is derived from biological experiments \cite{Elati07}.
% is assumed to be non-zero with probability $E$, which corresponds to the deregulation score to compute.
}\label{fig:EM}
\end{figure}

More precisely, according to the hLICORN model, each gene $g$ is connected with a set of co-regulated TFs split into a group of co-activators $\mathcal{A}$ and co-inhibitors $\mathcal{I}$.
 A binary deregulation variable $D_g$, assumed to be non-zero with probability $Y$, is then introduced to compare the true status $\mathcal{S}_g$ (under/over-expressed or normal) of each target gene in each tumor sample with its expected value $\mathcal{S}_g^*$, resulting from a truth Table (see Figure 
 %\ref{table:truth}
  \ref{fig:EM} (b)) and the inferred GRN. To avoid discretization of the data, the status of all genes are considered as hidden variables. 
%As the likelihood of the model is intractable due to the
%large number of hidden variables, the unknown model parameters (including the deregulation score E) are estimated using an
%EM-algorithm. 
The model is described in Figure \ref{fig:EM} (a).

%\newcolumntype{C}{>{\centering\arraybackslash}p{2em}}

%We make the follwing assumptions on the model:
%\begin{itemize}
%\item the expression $X_g$ of each gene $g$ follows a normal distribution, whose parameters depend on the hidden status $\mathcal{S}_g$ of $g$:
%$$X_g | \mathcal{S}_g=s \sim \mathcal{N}(\mu_s,\sigma_s), \ \ \mbox{with $\mu_s$ and $\sigma_s$ depending on $s$},$$
%\item the hidden status $\mathcal{S}_g$ of each TF follows an independant multinomial distribution with parameters $\alpha=(\alpha_-,\alpha_0,\alpha_+)$.
%\end{itemize}
As the number of hidden variables grows exponentially with the number of genes, the likelihood of the model rapidly becomes intractable. The unknown parameters, including
 %the means and the standard deviations of the Gaussians, the vector $\alpha$ and 
the deregulation score $Y$, are thus estimated using a dedicated EM-algorithm (see \cite{deregScore} for more details).
%We do not present it here but the interested reader can refer to \cite{deregScore} for more details. 
Note that the deregulation score $Y$ does not capture information about differentially expressed genes but genes whose expression does not correspond to the level expected from its regulator expression. 

%Consider for instance a gene $g$ regulated by a single TF $a$ activating $g$. If both $g$ and $a$ have a fold change of $10$, $g$ is differentially expressed but is not deregulated as the regulation relationship is conserved. Conversely, if the respective fold-changes of $g$ and $a$ are of $1$ and $10$, $g$ is not differentially expressed but is deregulated.   

%------------------------------------------------------------------------------
\subsection{Step 3: Identifying Deregulated TFs
% involved in the deregulation of the target genes
}\label{sec-Beta}
Step 3 consists in identifying TFs that cause deregulations of target genes. Our approach is based on linear regression models, in which we try to explain the deregulation score of all target genes in one sample (Step 2) using their co-regulator TFs as explanatory variables (Step 1).
Assume that we have
%the total number of genes $p$ is split into 
$q$ TFs and 
%$p-q$
$p$ target genes. Denote by $Y_{ij}$ the deregulation score of target gene $i$ ($1\leq i \leq p$) in sample $j$ ($1\leq j \leq n$) and $G:=(G_{i\ell})_{1\leq i \leq p,1\leq \ell\leq q}$ the GRN adjacency matrix, whose non-zero elements encode the structure (edges) of the graph.
We then cast our model as follows: 
\begin{equation}\label{eq:B}
\forall j \in \llbracket 1,n \rrbracket, \forall i \in \llbracket 1,p \rrbracket, Y_{ij}=G_{i\ell} \cdot B_{\ell j } + \varepsilon_{ij},
\end{equation}
or, in a matrix form, $Y=G\cdot B +\varepsilon$, where each element $B_{\ell j}$ of matrix $B$, to estimate, measures the deregulation importance of TF $\ell$ in sample $j$ and $\varepsilon$ stands for the presence of noise.
% in the model.

Solving the $B$-estimation problem (\ref{eq:B}) can be viewed as a classical multi-task linear learning problem, in which the 
%, which is known to be particularly critical in the high-dimensional setting. Note however that we are far from such a case,
%the 
number of observations
is 
%, which corresponds to 
the number of target genes $p$, 
%being extremely large compared to
the number of linear tasks is $n$ 
%(number of linear tasks) 
and 
%of the same order as 
the number of variables $q$.
% (number of variables).
To estimate $B$, we use a constrained least squares estimation procedure. As we only expect to find TFs positively causing the deregulation of their targets in each sample, we consider the induced constrained optimization problem:
\begin{eqnarray}
\forall j \in \llbracket 1,n \rrbracket, \ \  \hat{B}_{\cdot j} &:=& \underset{\beta\in \mathbb{R}^q}{\operatorname{argmin}} \Vert Y_{\cdot j} -G \beta \Vert_2^2, \label{eq:opt}\\
& s.t&  \ \ \forall \ell \in \llbracket 1,q \rrbracket, 0\leq \beta_{\ell} \leq 1 \nonumber
%\ \ \begin{cases} \forall \ell \in \llbracket 1,q \rrbracket, \beta_{\ell} \geq 0 &\\
% \sum\limits_{\ell = 1}^q \beta_q =1&
%\end{cases}\nonumber
\end{eqnarray}
where $\Vert . \Vert_2^2$ stands for the euclidian norm.
%The first constraint makes all coefficients of $\hat{B}$ positive, whereas the second constraint allows us to interpret $(\hat{B}_{\cdot j})_{1\leq j \leq n}$ as an influence deregulation score of TFs in each sample $j$. 
The closer $\hat{B}_{\ell j}$ is to 1, the more important the role of TF $\ell$ in the deregulation of its targets in sample $j$.
To solve Eq. (\ref{eq:opt}), we use the {\tt limSolve} R-package.

%\textcolor{red}{Etienne? + Add renormalisation de la matrice d'adjacence.}

%------------------------------------------------------------------------------
\subsection{Correcting Expression Data 
%by copy number correction
} \label{sec-correc}
Gene expression is commonly affected by copy number alterations (CNA) \cite{Aldred05}. Step 2 of our procedure is particularly sensitive to CNA, associating high deregulation scores to amplified or deleted  target genes \cite{deregScore}. Indeed, the number of copies of a gene can strongly influence its expression, independently from its regulators expression, making some regulations wrongly deregulated.

To remove CNA effects on gene expression and improve the rest of our analysis, we preprocess target genes expression data beforehand as proposed in \cite{SegCorr}. Gene expression is considered as linearly modified by CNA through the linear regression model:
\begin{equation}\label{eq:CNVcorrec}
X_{ij} = \alpha_0 + \alpha_1 \mbox{CNA}_{ij} + \varepsilon_{ij},
\end{equation}
where $X_{ij}$ is the expression of gene $j$ in sample $i$ and $\mbox{CNA}_{ij}$ its associated copy number. Let $\hat{\alpha}_0$ and $\hat{\alpha}_1$ be the estimated solutions of Eq. (\ref{eq:CNVcorrec}), the corrected expression is then given by:
$$\tilde{X}_{ij} = X_{ij} - \hat{\alpha}_0 - \hat{\alpha}_1 \mbox{CNA}_{ij}.$$

%\textcolor{red}{Note that we only correct target genes expression since a modification of a TF's expression by CNA directly affects its regulators expression, unless the regulation goes wrong.}

%with $\hat{\alpha}_0$ and $\hat{\alpha}_1$ estimated solutions of Equation (\ref{eq:CNVcorrec}). 
%
%Note that instead of applying this correction on the expression of all genes, we only selected the target genes. Indeed, when a CNV alteration occurs on the upper layer of the model, not only the expression of the altered TF is modified, but also the expression of all its non-deregulated targets. 

\section{Results and Discussion}
\subsection{The Bladder Cancer Data Set}\label{sec-data}
We apply our method on bladder cancer data, produced in the framework of the 
%Cartes d'Identit\'e des Tumeurs (CIT) French national program and freely available in the R-package {\tt CoRegNet}.
Cancer Genome Atlas (TCGA) project and available at the Genomic Data Commons Data Portal (\url{https://portal.gdc.cancer.gov/}). 
These data include a set of 
401
%171
bladder cancer samples with gene expression and copy number for a total number of 
15,430 genes,
%16,495 genes,
 split into 
2,020 
%1,693
 TFs and 
13,410 
%14,802
 targets.
Gene expression data were produced using RNA-sequencing on bladder cancer tissues. Preprocessing is done by log-transformation and quantile-normalization of the arrays. 
Missing values are estimated using nearest neighbor averaging \cite{Troyanskaya01}.
TCGA samples are analyzed in batches and significant batch effects are observed based on a one-way analysis of variance in most data modes. We apply Combat \cite{Johnson07} to adjust for these effects. 
%Note that although this method was originally designed for microarray gene expression data, it was recently shown to be suitable for RNA-sequencing too \cite{Conesa16}.
Genes are finally filtered based on their variability: 
%\cite{Hackstadt09}: 
among them, 
%15,430
%%16,495
%Among them, 
we only keep the $75\%$ most varying genes. 
 %The filtered data set we used thus contained 
%10,485 
%12,371
% genes, including 
%1,075 
%1,264
% TFs. 

Based on RNA-seq data analysis from the TCGA data portal, samples are split into five subtypes% \citep{TCGA_subtypes}
: basal-squamous (BaSq), luminal (Lum), luminal-infiltrated (LumI), luminal-papillary (LumP) and neuronal (NE) with different characteristics \cite{TCGA_subtypes} (Table \ref{samples}). 
%For more details on these subtypes, the interested reader can refer to \cite{TCGA_subtypes}.
%The three luminal subtypes present similar characteristics (high expression of urothelial differential markers, e.g. GATA3, PPARG, FOXA1) but also differences in purity and expression profiles when comparing to p53, EMT and stromal gene signatures. 
%The basal-squamous subtype is associated with high-expression of basal, stem-like (CD44, KRT5, KRT6A, KRT14) and squamous differentiation markers (TGM1, DSC3, PI3).
%The neuronal subtype is finally closely related to high expression of neuronal differentiation and development genes. 

%A large number of bladder samples subgroups exist in the litterature, with length from $2$ to $10$ and sharing some characteristics. In this paper, we use the very new classification from \citep{Newclassif}, which arised from the international consortium of MIBC molecular classification.
%To complete...

\begin{table}[!ht]

\begin{center}

%\begin{tabular}{c c}
%Subtype & Number of samples \\
%\hline 
%Luminal-papillary & 134 samples \\
%Luminal-infiltrated & 74 samples \\
%Luminal & 44 samples \\
%Basal-squamous & 131 samples \\
%Neuronal & 18 samples
%%Basal squamous& 152\\
%%Luminal Non specified & 21\\
%%Luminal Papillary & 127\\
%%Luminal Unstable & 53\\
%%Neuroendocrine-like & 6\\
%%Stroma-rich & 44
%\end{tabular}

 \begin{tabularx}{0.55\textwidth}{c c c c c c}
\toprule
Subtypes &BaSq & Lum & LumI & LumP & NE \\
\cmidrule(lr){1-1} \cmidrule(lr){2-2} \cmidrule(lr){3-3} \cmidrule(lr){4-4}\cmidrule(lr){5-5}\cmidrule(lr){6-6}
Samples &131&44&74&134&18\\
\bottomrule
\end{tabularx}

\end{center}
\caption{Molecular subtypes distribution of the 401 bladder cancer samples
%The five subtypes of the bladder cancer data set, established by RNA-seq analysis from the TCGA data portal 
\cite{TCGA_subtypes} 
%and their samples distribution
.}\label{samples}
\end{table}

%related to luminal A breast cancer tumors, \textit{basal}, to basal-like breast cancer tumors and \textit{p53}, which differs from the luminal subtype by an activated wild-type p53 gene expression signature.
%Previous analysis of RNA-seq data from the TCGA data portal led to the identification of four sample clusters \citep{TCGA2}:
%\begin{itemize}
%\item cluster I (``papillary-like''), enriched in tumors with papillary morphology,
%\item cluster II (``luminal''), sharing characteristics with luminal A breast cancer tumors when combined with cluster I,
%\item cluster III (``basal-like''), similar to basal-like breast tumors,
%\item cluster IV, with nothing clearly established.
%\end{itemize}
%This analysis was performed on the TCGA gene expression data based on a bootstrapped ensemble clustering algorithm that merges the output of hierarchical and k-means clustering. More recently, using a centroid-based predictor, \citet{Biton14} established a classification of the 171 samples from the CIT data set into the same four classes: ``papillary-like'' (83), ``luminal'' (49), ``basal-like'' (22) and ``cluster IV'' (17).

\subsection{Description of the Procedure Results}
\paragraph{GRN network.}
To validate our method, we have to provide a tissue-specific reference GRN (Step 1), which is computed given a first set of reference samples.
%In this work, we used as a reference the whole TCGA data set presented in Section \ref{sec-data}. 
%We thus have to choose a set of samples on which no deregulation score will be computed.
 In many cancers, the pure normal tissue of origin is not available. 
 %In this work, we used as a reference 
 Here, we work with the five different subtypes of the TCGA data set presented in Section \ref{sec-data}.
 % and corresponding to the five subtypes of bladder cancer.
Using samples from one subtype as test cases and the rest as reference, we infer five different GRNs. 
Each of them reflect
%The inferred regulatory network will thus reflect
 averaged relationships between genes for patients
 % suffering of bladder cancer
 who are not part of one specific subtype. 
 %However, 
 Due to the very-high heterogeneity of cancers, especially of bladder cancers \cite{Knowles14,Togneri16}, we think that our method will still point to relevant deregulations of specific subtypes.

After calibrating the internal parameters of the hLICORN agorithm, the GRNs we infer are made of an averaged total number of $28,246$ edges connecting $586$ TFs to $3,432$ of their targets. 
%Among these $3,648$ targets, $472$ are TFs and also regulate other genes.
These networks are relatively sparse, each of the target genes being associated with an averaged number of 8 TFs. 
%The remaining genes, which are not connected to any other genes, were removed from the rest of the analysis. 

%\begin{figure}[!ht]
%\begin{center}
%\includegraphics[width=10cm]{graph.png}
%\end{center}
%\caption{Gene regulatory network inferred by the hLicorn algorithm using gene expression data from a set of bladder tumor samples.}\label{fig:net}
%\end{figure}

\paragraph{Deregulation scores.}
We then run five times the EM procedure (Step 2) on the five subsets of the gene expression data matrix to compute a deregulation score of each target gene in each sample of each subtype.
 From now on, all samples are treated individually, the results reflecting how genes behaved in each sample of one subtype in comparison to reference samples from all other subtypes. %As can be seen in Figure \ref{fig:scores}, the deregulation scores appear to be larger than $0.5$ in approximately 9,000 cases. 
%As expected, most of the target genes are not deregulated at all.
%(see Figure \ref{fig:scores}). 
%However, a deregulation score larger than 0.95 is observed in around 768 cases. 
 
%%
%\begin{figure}[!ht]
%\begin{center}
%\includegraphics[width=10cm]{hist_dereg.jpeg}
%\caption{Histogram representing the deregulation scores distribution across all samples and all target genes. The y-axis is log-scaled.}\label{fig:scores}
%\end{center}
%\end{figure}

To check the effect of the copy number correction we apply at the beginning of our procedure,
% we ran it twice, with or without correction. 
we 
%then 
compare the distribution of the deregulation scores across copy number states.
%with CNV thresholded data
To this aim, we use the TCGA CNA thresholded data set, which 
%. This data set 
associates to each gene-sample pair a copy number state of ``0'' for the diploid state (two copies), ``1'' for a copy number gain, ``-1'' for a copy number loss, ``2'' for an amplification and ``-2'' for a deletion.
We then test for significant differences between the diploid state and the altered states (-2,-1,1,2) using Student tests. Results in terms of p-values, which are corrected for multiple hypothesis testing using the FDR \cite{Benjamini95}, are presented in Table \ref{tab:CNV}. With corrected p-values ranging from 0.10 to 1, deregulation scores are no longer associated with CNA.
% with p-values ranging from 0.096 to 1.
%Figure \ref{fig:compCNV} shows a comparison between deregulation scores computed with corrected and non-corrected data.
%Figure \ref{fig:cumfig} represents the empirical cumulative distribution function of the deregulation scores using the non-corrected gene expression data.
% As expected, contrary to the results presented in \citet{deregScore},
% scores are yet not associated with CNV alterations with $p$-values for Student tests ranging from 0.07 to 0.88 (Table \ref{tab:CNV}). 
 
 \begin{table}[!ht]
 \begin{center}
 {\small
  \begin{tabularx}{1\textwidth}{c c c c c c c c c c c c c c c c c c c c}
  \toprule
% CNV alterations & -2 & -1 & 1 &2\\
% p-values & 0.51 & 0.88 & 0.31 & 0.072 
\multicolumn{20}{c}{ Subtypes} \\
\cmidrule(lr){1-20}
\multicolumn{4}{c}{BaSq} & \multicolumn{4}{c}{Lum} &\multicolumn{4}{c}{LumI} & \multicolumn{4}{c}{LumP} &\multicolumn{4}{c}{NE}\\
\cmidrule(lr){1-4}  \cmidrule(lr){5-8}  \cmidrule(lr){9-12} \cmidrule(lr){13-16}\cmidrule(lr){17-20}
-2&-1&1&2 & -2&-1&1&2 &-2&-1&1&2  &-2&-1&1&2  & -2&-1&1&2 \\
%\cmidrule(lr){2-5}
%Subtypes& -2&-1&1&2\\
%\cmidrule(lr){1-1} \cmidrule(lr){2-5} 
1&0.81&0.28&1 &1&1&1&1&1&0.25&0.10&0.60&1&1&1&1&1&1&1&1\\
\bottomrule
 \end{tabularx}
 }
 \end{center}
  \caption{Corrected p-values for Student tests when comparing the distribution of the deregulation scores between the diploid state
 %with no CNV alteration
  (0) and 
  each altered state (-2,-1,1,2) for each subtype.
  %amplifications/deletions (CNV=$\pm$ 1,2)
  }\label{tab:CNV}
 \end{table}

\paragraph{Deregulated TFs.}
We finally apply Step 3 of our method to identify TFs involved in the deregulation scores of the target genes, that is having a non-zero coefficient in $\hat{B}$, as given in Eq. (\ref{eq:B}). 
We then rank the TFs according to their number of non-zero coefficients across all samples belonging to each specific subtype. Results are presented in Table \ref{tab:b}.

\begin{table*}[!ht]
\begin{center}
{\small
 \begin{tabularx}{1\textwidth}{c c c c c c c c c c}
\toprule
\multicolumn{10}{c}{Subtypes}
\\
\midrule
 \multicolumn{2}{c}{BaSq}& \multicolumn{2}{c}{Lum} & \multicolumn{2}{c}{LumI} &\multicolumn{2}{c}{LumP} &\multicolumn{2}{c}{NE}
 %&\multicolumn{2}{c}{Stroma-rich}
 \\
\cmidrule(lr){1-2} \cmidrule(lr){3-4}\cmidrule(lr){5-6}\cmidrule(lr){7-8}\cmidrule(lr){9-10}
%\cmidrule(lr){11-12}
 TF & \small{$\%  \hat{B}$} & TF &\small{$\% \hat{B}$}  & TF & \small{$\% \hat{B}$} &TF & \small{$\% \hat{B}$} &TF & \small{$\% \hat{B}$} 
 %&TF&$\#\hat{B}$
 \\
\midrule
SPOCD1 &\small{92\%}& ZNF268 &91\% &TSHZ1 &88\% &RARB &84\%& FAIM3& 89\%\\
ZNF382 &\small{86\%}& HES2& 80\%& ZNF354B& 88\%& RFX5 &84\%& SMARCA2& 83\% \\
RCOR2 &\small{86\%}& TBX2& 80\%& AR &85\%& CBFA2T3& 83\%& RARB &78\%\\
ATM &\small{83\%}& PRDM8 &75\%& HES2& 82\%& TBX18& 81\%& ZNF235& 78\%\\
HABP4 &\small{83\%}& TSHZ1& 75\%& HTATIP2& 81\%& TBX3& 79\%& TBX2& 72\%\\
IRX3& \small{82\%} &ZNF354C &73\% &MAFG &80\%& PTRF& 79\% &STAT3 &72\% \\
IFI16 &\small{79\%}& RARB &70\%& ENO1 &80\% &TBX2 &70\%& HIF1A &72\%\\
TEAD2& \small{79\%} &KLF13 &70\%& TBX2& 74\%& PPARG &76\% &THRA &72\%\\
NOTCH4 &\small{79\%} &SCML2 &68\%& ZNF563&74\%& NCOR2 &75\%& PIR& 67\%\\
%ZNF211& 79\% &SMARCA2& 68\%& SOX7 &72\%& MSX2& 75\%& PRDM8 &67\%\\
SNAI2& \small{79\%} &SNAI3& 68\%& IRX3& 72\%& ZFP2& 75\%&FOSL1& 67\%\\
%LMO3 &76\%& HIVEP3 &68\%& MCM3 &72\%& MYOCD &74\%& RORC& 67\%\\
%MAFG &76\%& VGGL1 &68\%& SMARCD3 &70\%& SMARCD3 &72\%& NAP1L2& 67\%\\
%CRY1& 76\%& NKD2 &68\% &MYCN &70\% &PRI& 72\%& NR3C2 &67\%\\
%HOXC9 &75\%& MESP1 &66\% &SP4 &69\%& TCEAL1& 72\%& NCOA1 &67\%\\
%SPOCD1 & 93\% &SMARCD3 &81\%& RARA& 91\%& HES2& 87\% &HES2 &83\%& HES2 &90\%\\
%IRX3 &85\%& ANKS1A& 81\% &PIAS2 &86\% &SOX7 &83\%& TBX2& 83\%& SPOCD1 &90\%\\
%PEG3& 83\%& ZNF423& 81\%& CBFA2T3& 80\% &ZNF671& 79\%& MAFG &83\%& TBX2 &81\%\\
%SMARCD3 &79\%& NOTCH4 &76\%& SALL2& 78\%& SPOCD1& 79\%& PIR &83\%& SPIB& 77\% \\
%SNAI2& 79\%& HOXC9 &76\% &LHX6 &78\% &HOXB3 &77\%& IRX3& 83\%& IRX3& 75\%\\
%NAP1L2& 76\% &ZNF563 &76\%& RARB& 77\%& TBX2 &74\% &SETBP1 &83\% &ZNF713& 75\%\\
%ZSCAN12 &76\% &SPOCD1& 71\% &SOX7 &76\%& NR3C2 &74\%& RARB &83\%& MAFG &73\%\\
%ZNF433& 75\%& HES2 &71\% &BACH2 &76\%& BTBD11& 74\%& SOX15& 83\%& SMARCD3& 71\%\\
%JARID2& 75\%& IKBKB &71\% &SMARCD3& 74\% &PPARGC1B &74\%& MITF& 83\%& PER3& 71\%\\
%HOXA13 &73\%& HOXC6 &71\%& NFIC &74\% &TEAD4 &72\%& HTATIP2& 83\%& ZNF563& 71\%\\
\bottomrule
 \end{tabularx}
 }
\end{center}
\caption{List of the 10 most important TFs for explaining the deregulation scores of their targets and number of non-zero coefficients in $\hat{B}$ (in $\%$) across all samples from each subtype.}\label{tab:b}

\end{table*}

%Among them, to complete.

\subsection{Discussion}
\paragraph{Top TFs include biomarkers of bladder cancer.}
Among TFs of Table \ref{tab:b}, we retrieve characteristic genes of bladder cancer subtypes. 
For instance, SNAI2, which is deregulated across 79\% of the BaSq samples, is particularly well-known for its implication in EMT pathways for cancer patients \cite{Cobaleda07} and its capacity to discriminate between basal and luminal subgroups \cite{Mistry14}.
The presence of NOTCH4 in BaSq samples
%deregulated for 79\% of the basal samples,
 is particularly interesting as it is part of the NOTCH pathway, whose inactivation tends to promote bladder cancer progression \cite{Maraver15}. Research works also focus on 
 %the effect of the NOTCH pathway 
 its implication on the basal subgroup \cite{Greife14}.
Similarly, TBX2, involved in all three luminal subtypes is an indicator of luminal cancers  \cite{Dhawan15}. We can finally emphasize the presence of PPARG in LumP, whose high level of expression is used to describe luminal subtypes \cite{Choi14}.
%HEs2 que dire?

\paragraph{Deregulation is complementary to differential gene expression analysis}
Differential gene expression analysis consists in performing statistical analysis to discover quantitative changes in terms of expression levels between groups. It is frequently used in cancer research to identify genes with important changes between tumor and normal samples, called differentially expressed genes (DEGs) \cite{limma}.

We perform differential gene expression analysis using the \texttt{R}-package \texttt{limma} \cite{limma} on all samples from each subtype when comparing to samples from all other subtypes. 
We then verify whether the identified DEGs are different from the deregulated TFs derived from our method (Figure \ref{fig:DEG}). To this aim, we use the following thresholds: a gene is called DEG for p-values smaller than 0.01 whereas it is deregulated for a subtype as soon as it is deregulated ($\hat{B}\neq 0$) for more than the 50\% of the subtype samples.
This threshold is purely arbitrary but is not crucial, as the results remain almost the same with slight changes.
As shown in Figure \ref{fig:DEG},
except for BaSq and LumP subtypes, more than the 70\% of the identified deregulated TFs are not differentially expressed, which means that our procedure does not only point to DEGs. 
%However, it may be used complementary to differential gene expression analysis to discover biomarkers of cancer subtypes.

\def\firstcircle{(0,0) circle (0.8cm)}
\def\thirdcircle{(0:1cm) circle (0.8cm)}

\begin{figure}[!ht]
\begin{center}
\begin{tikzpicture}[scale=1]
        \begin{scope}[fill opacity=0.5]
        \fill[magenta] \firstcircle;
        \fill[cyan] \thirdcircle;
            \end{scope}
             \begin{scope}
        \draw \firstcircle node[left] {$879$};
        \draw \thirdcircle node [right] {$48$};
               \draw(0.5,0) node {$107$};
                       \draw(0.5,1.2) node {BaSq};
    \end{scope}
    
            \begin{scope}[shift={(2.8cm,0cm)}, fill opacity=0.5]
           \fill[magenta] \firstcircle;
        \fill[cyan] \thirdcircle;
            \end{scope}
                        \begin{scope}[shift={(2.8cm,0cm)}]
        \draw \firstcircle node[left] {$193$};
        \draw \thirdcircle node [right] {$65$};
         \draw(0.5,0) node {$17$};
          \draw(0.5,1.2) node {Lum};
    \end{scope}
    
            \begin{scope}[shift={(5.6cm,0cm)}, fill opacity=0.5]
        \fill[magenta] \firstcircle;
        \fill[cyan] \thirdcircle;
           \end{scope}
            \begin{scope}[shift={(5.6cm,0cm)}]
        \draw \firstcircle node[left] {$366$};
        \draw \thirdcircle node [right] {$57$};
         \draw(0.5,0) node {$23$};
          \draw(0.5,1.2) node {LumI};
    \end{scope}
    
            \begin{scope}[shift={(8.4cm,0cm)}, fill opacity=0.5]
         \fill[magenta] \firstcircle;
        \fill[cyan] \thirdcircle;
      \end{scope}      
          \begin{scope}[shift={(8.4cm,0cm)}]
        \draw \firstcircle node[left] {$784$};
        \draw \thirdcircle node [right] {$45$};
         \draw(0.5,0) node {$91$};
          \draw(0.5,1.2) node {LumP};
    \end{scope}
    
            \begin{scope}[shift={(11.2cm,0cm)}, fill opacity=0.5]
        \fill[magenta] \firstcircle;
        \fill[cyan] \thirdcircle;
     \end{scope}
            
          \begin{scope}[shift={(11.2cm,0cm)}]
        \draw \firstcircle node[left] {\textcolor{black}{337}};
        \draw \thirdcircle node [right] {$46$};
         \draw(0.5,0) node {$12$};
          \draw(0.5,1.2) node {NE};
    \end{scope}
\end{tikzpicture}
\end{center}
\caption{Venn Diagrams representing the number of DEGs (in pink), the number of deregulated TFs identified by our method (in blue) and their intersection.}\label{fig:DEG}
\end{figure}
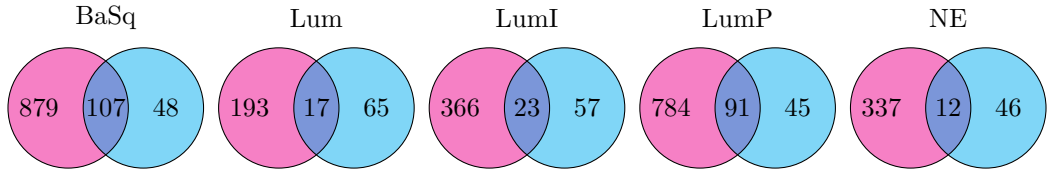

\section*{Conclusion}
With the aim of understanding the deregulation processes in tumoral cells, we develop a 3-steps strategy that measures the influence of TFs in the deregulation of genes in tumor samples.
A list of TFs characterizing given subtypes can then be established. Even if a biological experimental validation
%of such lists 
should be done in future work, it seems that it can be used complementary to differential gene expression analysis to point to potential biomarkers of cancers. 

An open question, which has also to be tackled, is to determine in which extend the information carried by mutations can explain the deregulations. Mutation data are particularly hard to explore in this context due to various reasons : first of all, mutations do not necessarily affect gene expression. Secondly, in cancers, besides the most significant mutated genes, many sequencing projects have shown that genes are mutated in less than 5\% of the samples.

In this work, among the identified TFs of Table \ref{tab:b}, we find ATM, which is highly deregulated (83\%) and mutated (15\%) for BaSq samples. Mutations of ATM have been recently shown to be associated with shorter survival in urothelial cancers \cite{Yin18}.
% making it an interesting target. 
As a preliminary result, we observe that 95\% of the mutated BaSq samples corresponds to non-zero $\hat{B}$ coefficients (Table \ref{tab:mut}). 
This table is unfortunately still too unbalanced to positively conclude for a significant association but supplementary works need to be done to go further.

 \begin{table}[!ht]
 \begin{center}
  \begin{tabularx}{0.36\textwidth}{c c c}
  \toprule
% CNV alterations & -2 & -1 & 1 &2\\
% p-values & 0.51 & 0.88 & 0.31 & 0.072 
& $\hat{B}\neq 0$ & $\hat{B}=0$ \\
\cmidrule(lr){2-3}
Non mutated& 91 & 21\\
Mutated &18 & 1 \\
\bottomrule
 \end{tabularx}
 \end{center}
  \caption{Confusion matrix indicating the association between mutation and deregulation status $\hat{B}$ for TF ATM across all 131 basal samples.
  %amplifications/deletions (CNV=$\pm$ 1,2)
  }\label{tab:mut}

 \end{table}

\label{sect:bib}
\bibliographystyle{plain}
\bibliography{refs2}

%------------------------------------------------------------------------------

%------------------------------------------------------------------------------
\end{document}